# Three-dimensional Near-field Analysis Through Peak Force Scattering-type Near-field Optical Microscopy

Haomin Wang[1], Jiahan Li[2], James H. Edgar,[2] and Xiaoji G. Xu[1]*

[1]Department of Chemistry, Lehigh University, Bethlehem, PA 18015, USA
[2]Tim Taylor Department of Chemical Engineering, Kansas State University, Manhattan, KS 66506, USA

*Email: xgx214@lehigh.edu

**Abstract**

Scattering-type scanning near-field optical microscopy (s-SNOM) is instrumental in exploring polaritonic behaviors of two-dimensional (2D) materials at the nanoscale. A sharp s-SNOM tip couples momenta into 2D materials through phase matching to excite phonon polaritons, which manifest as nanoscale interference fringes in raster images. However, s-SNOM lacks the ability to detect the progression of near-field property along the perpendicular axis to the surface. Here, we perform near-field analysis of a micro-disk and a reflective edge made of isotopically pure hexagonal boron nitride ($h$-$^{11}$BN), by using three-dimensional near-field response cubes obtained by peak force scattering-type near-field optical microscopy (PF-SNOM). Momentum quantization of polaritons from the confinement of the circular structure is revealed *in situ*. Moreover, tip-sample distance is found to be capable of fine-tuning the momentum of polaritons and modifying the superposition of quantized polaritonic modes. The PF-SNOM-based three-dimensional near-field analysis provides detailed characterization capability with a high spatial resolution to fully map three-dimensional near-fields of nano-photonics and polaritonic structures.

**Introduction**

Phonon polaritons (PhPs), quasi particles from a hybrid of electromagnetic (EM) wave and collective phonon oscillations, have been a research focus of nanophotonics. Utilization of PhPs holds the promise for bridging electronic and photonic technologies in infrared frequencies.[1] PhPs propagate along the interface of the materials with opposite signs of the real part of the dielectric function,[2] or in the volume of hyperbolic materials.[3] PhPs have much more compressed wavelength than free-propagating photons, leading to a locally enhanced EM field. Recent advances of two-dimensional (2D) materials provide a range of new polariton-supporting materials.[1, 4, 5] Hexagonal boron nitride ($h$-BN) supports surface phonon polaritons and hyperbolic phonon polaritons and has been a fundamental material for the development of nanotechnology based on polaritons.[6-11] Nano or micro resonators made of polaritonic materials are building units for next-generation polaritonic devices, and it becomes important to understand their near-field behaviors in the vertical direction, since resonators are usually integrated within a heterostructure, where near fields of resonators may interfere with other components.

The short wavelength of the polaritons is associated with high momentum, which cannot be directly coupled from free-space propagating photons due to the restriction from phase matching.[12] Scattering-type scanning near-field optical microscopy (s-SNOM) is a popular scanning probe microscopy technique that can both excite and detect PhPs. The metallic atomic force microscope (AFM) tip locally enhances EM field at its apex with high spatial frequencies and is capable of effectively exciting polaritons in 2D materials.[12-16] The excitation of polaritons modifies the scattering properties of the AFM tip, leading to changes of amplitude and phase of the scattered light, but not in frequency. To detect the near-field signal from the far-field background of same optical frequency, the AFM cantilever is externally driven to oscillate at its mechanical resonance in tapping mode AFM. The scattered light from the AFM tip is optically detected and demodulated by a lock-in amplifier at a non-fundamental harmonic of the cantilever oscillation frequency, from which one can extract the s-SNOM signal.

The operation mechanism of s-SNOM demarcates its limitation: the near-field generated from s-SNOM is an average of near-field responses at many different tip-sample distances during the tip oscillation. Different tip-sample distances correspond to different spatial confinement of EM field underneath the AFM tip, thus leading to different excitation conditions (Figs. 1a-b). However, s-SNOM only provides one-value signal associated with a demodulation order, thus only yielding a near-field image in two lateral dimensions, rather than providing three-dimensional (3D) near-field responses with both lateral and vertical Cartesian coordinates. More information about polaritons in photonic nanostructures can be obtained, if all three dimensions of near-field responses are available. Here, we utilize a new type of near-field microscopy, the peak force scattering-type near-field optical microscopy (PF-SNOM),[17] to obtain a 3D near-field response cube of polaritonic isotopically pure $h$-$^{11}$BN microstructures. We perform near-field analysis along the tomographic and sectional profiles of the 3D response cube at a series of infrared frequencies, revealing subtleties of phonon polaritons through correlations on spatial and spectral features that were not previously available for s-SNOM.

**Method for Collection of 3D near-field Response Cube**

PF-SNOM operates with the peak force tapping (PFT) mode of AFM. The sample is oscillated vertically at a low frequency (~4 kHz) that is much below the resonant frequency of the AFM cantilever. Under PFT mode, instantaneous tip-sample distance is deterministic from the extension of the piezoelectric sample stage and the dynamic deflection of the AFM cantilever. PF-SNOM simultaneously measures the scattered light signal from tip-sample region and the instantaneous tip-sample distance (Fig. 1c). The relation of the tip-sample distance versus raw scattering signal from the tip is obtained through a correlation of signal waveforms in the time domain (Fig. 1d). Because the oscillation amplitude of the tip-sample distance in PFT mode spans more than a hundred nanometers, a linear far-field background fit can be performed on the region of large tip-sample distances where the near-field signal is negligible. The resulting linear background is back extrapolated to all tip-sample distances and removed from the raw scattering signal. The result is a vertical near-field interaction curve with an explicit tip-sample distance dependence (Fig. 1e).

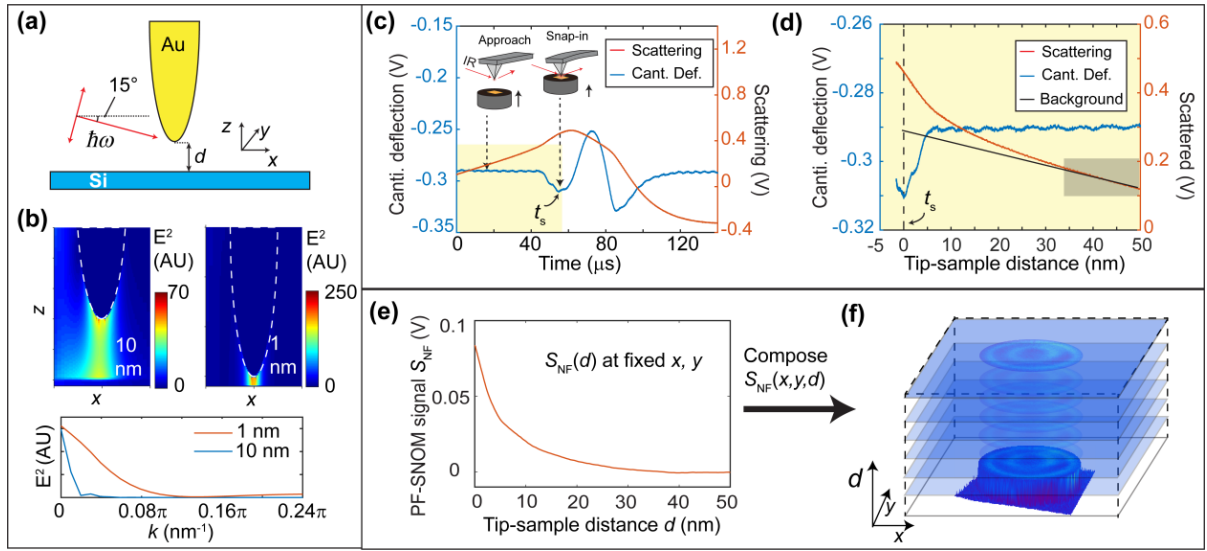

**Figure 1.** FDTD simulation on the effect of tip-sample distance and explicit tip-sample distance information in PF-SNOM. (a) The scheme used in FDTD simulation. A gold cone with end radius of 30 nm is used to mimic AFM tip, and the light illumination angle is 15°. (b) FDTD simulations of EM field enhancements with $d = 10$ and 1 nm under 1403 cm$^{-1}$ illumination. FFTs of signal profile along $x$-axis in the angular momentum $k$-space are plotted in the lower panel. Shorter tip-sample distance couples with higher momenta. (c) Simultaneously obtained cantilever deflection and scattered light from the tip-sample region during one cycle of peak force tapping. Insets schematically illustrates the tip-sample distance during PFT. The PF-SNOM sampling region is right before the snap-in time point $t_s$. (d) Cantilever deflection curve, scattered light and linearly fitted background shown in the tip-sample distance domain at the yellow region in (c). $t_s$ corresponds to $d = 0$ which is indicated by a vertical dashed line. The grey shadow indicates the background fitting range. Data in (c-d) are obtained on the Au surface at 1600 cm$^{-1}$. (e) A typical vertical near-field interaction curve. (f) A 3D near-field response cube is formed by assembling vertical interaction curves.

Vertical near-field interaction curves at every lateral scanned position are then assembled to form a 3D near-field response cube. The 3D near-field response cube $S_{\text{NF}}(x, y, d)$ includes near-field signals $S_{\text{NF}}$ at each lateral scanned position $(x, y)$, as well as each tip-sample distance $d$ (**Fig. 1f**). We collect many 3D near-field response cubes at a series of infrared frequencies. In the near-field tomography, the collected 3D near-field response cube is analyzed on the tomographic (with fixed $d$) and sectional (with fixed $x$ or $y$) directions to reveal correlations between the spatial frequency and the excitation frequency.

**Results**

**Tomographic Analysis of Near-field Responses**

We investigate a circular micro-disk and a reflective edge made of the isotopically pure hexagonal boron nitride ($h$-$^{11}$BN).[18] Hexagonal boron nitride is a two-dimensional van der Waals material that supports hyperbolic phonon polaritons (HPhPs).[3, 19-26] Isotopically pure $h$-$^{11}$BN is advantageous over regular $h$-BN in that it has smaller damping from the phonon relaxation, which leads to a longer polariton propagation length.[24] On the other hand, circular optical nano-resonators are known for their symmetrical fringe patterns and high quality factors when on resonance.[27-30] The investigated $h$-$^{11}$BN micro-disk is 9 μm in diameter and 70 nm in thickness as revealed by the AFM topography image (Fig. 2a). 3D near-field response cubes are collected with PF-SNOM at a series of infrared frequencies. Tomographic $x$-$y$ near-field images from response cubes at three tip-sample distances ($d$) and three frequencies ($\omega$) are displayed in Figs. 2b-d.

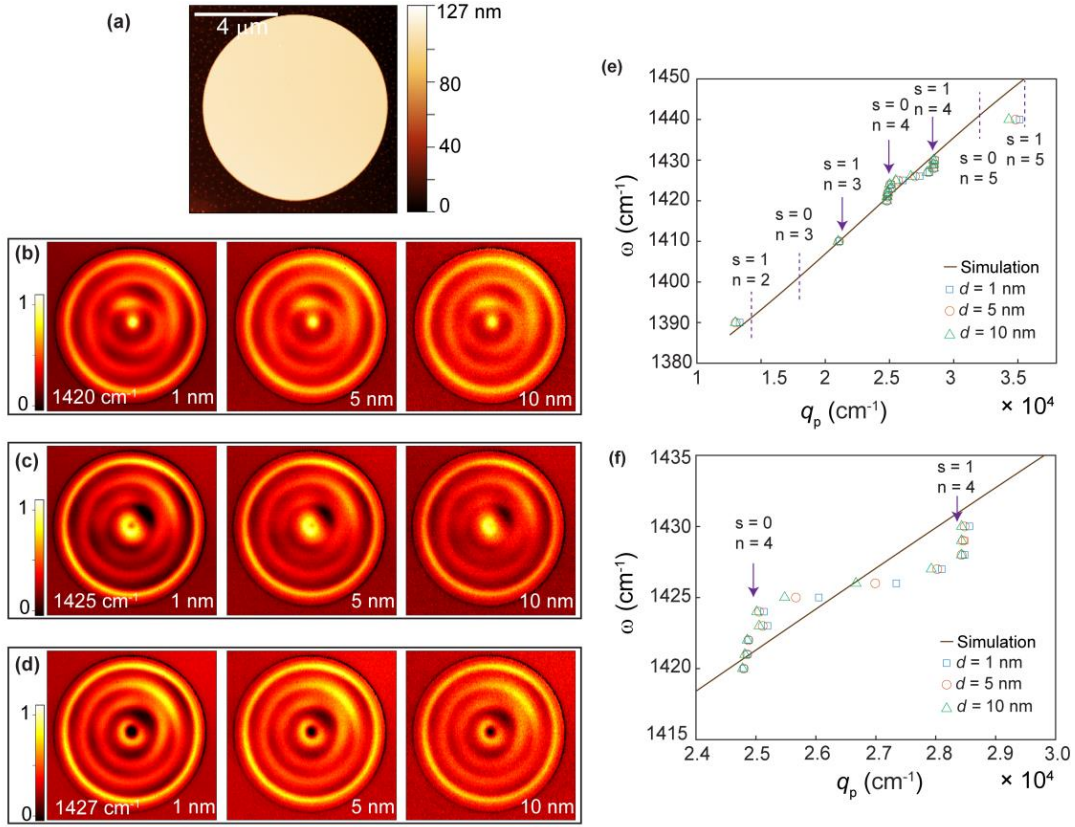

**Figure 2.** Fringe patterns, vertical near-field responses and dispersion relation of the $h$-$^{11}$BN disk through near-field analysis. (a) AFM topography of an $h$-$^{11}$BN micro-disk with a diameter of 9 μm and a thickness of 70 nm, the scale bar is 4 μm. (b-d) Normalized tomographic PF-SNOM images at tip-sample distances $d = 1, 5$ and 10 nm (in column) and at $\omega = 1420, 1425$ and 1427 cm$^{-1}$ (in row). In (c) and (d), positions of fringes shift inward as $d$ increases. (e) The $\omega - q_\mathrm{p}$ dispersion relation of the $h$-$^{11}$BN micro-disk extracted from PF-SNOM results. Experimental data (blue squares, red circles, and green triangles) are overlaid with calculated dispersion relation (brown curve). Calculated $k_{sn}$ correspond to three observed resonant conditions are shown as purple arrows, others are shown as dashed purple lines. Mode numbers $(s, n)$ are displayed along with resonant conditions. Details in the 1420-1430 cm$^{-1}$ range are shown in (f).

At first glance, circular fringes of the tomographic images from PF-SNOM resemble what one would obtain from regular s-SNOM with the high order lock-in demodulation (see Supplementary Fig. S1). However, tomographic images at different tip-sample distances $d$ reveal additional details that are missing from regular s-SNOM images. As $d$ is increased from 1 to 10 nm, the spatial patterns of the tomographic images evolve and are dependent on the infrared frequency $\omega$. For $\omega = 1420$ cm$^{-1}$, the fringe patterns do not exhibit noticeable changes. However, fringe patterns at $\omega = 1425$ and 1427 cm$^{-1}$ exhibit discernable changes: as $d$ increases, the central hole disappears or shrinks, indicating a change of spatial frequency of PhPs. An animation of the progression of the near-field images at different tip-sample distances is included in the Supplementary Movie S1.

We then extract the periodicity of the circular fringes of the $h$-$^{11}$BN disk to obtain the energy ($\omega$) versus momentum ($q_P$) dispersion relations (Fig. 2e-f). Details of the extracting method are described in Supplementary Fig. S2. The experimental dispersion relation exhibits an interesting behavior: the polariton momentum exhibits a fixed value at $2.5 \times 10^4$ and $2.85 \times 10^4$ cm$^{-1}$ over a small range of infrared frequency regardless of the change of tip-sample distance $d$; between these two fixed values, the momentum of polariton changes with $d$; larger $d$ corresponds to lower spatial frequencies of the polariton at a given excitation energy. The dispersion relation of the $h$-$^{11}$BN disk clearly deviates from the theoretical prediction of $h$-$^{11}$BN film that does not consider the geometrical confinement.[3] To confirm this is a general behavior of circular micro-disk of $h$-$^{11}$BN, we performed PF-SNOM measurement on another circular disk of different thickness, a similar behavior was found (see Supplementary Fig. S3).

We hypothesize that the confinement of the circular geometry leads to the quantized spatial frequencies in the lateral dimension. A question arises: Do polaritons behave similarly in the case of an $h$-$^{11}$BN edge that lacks symmetry for the geometrical confinement? To answer this question, an edge of an $h$-$^{11}$BN flake was measured by PF-SNOM (Fig. 3). The thickness of the flake is of 40 nm. Tomographic PF-SNOM images at 1415 cm$^{-1}$ from the 3D near-field response cube at three different tip-sample distances $d$ are shown in Fig. 3b-d. Close examination of the fringe spacing after averaging (Fig. 3e) reveals a small change of the periodicity: as tip-sample distance $d$ increase, the polariton wavelength $\lambda_p$ also slightly increases. The extracted dispersion relation of the polaritons in the $h$-$^{11}$BN flake at different tip-sample distances and light frequencies is displayed in Fig. 3f. In contrast to the behavior of the micro-disk, the momentum of polariton in the BN flake does not exhibit fixed values. Also, the momentum can be slightly tuned by the tip-sample distance $d$: longer $d$ corresponds to a smaller momentum of the polaritons. An animation of the progression of the near-field images of the $h$-$^{11}$BN edge at different tip-sample distances is included in Supplementary Movie S2.

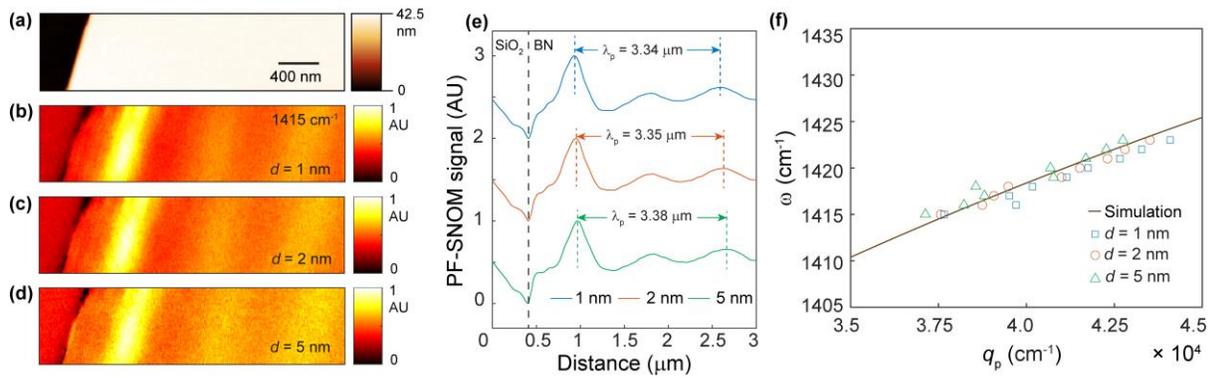

**Figure 3.** Near-field analysis on the edge of an $h$-$^{11}$BN flake. (a) Topography of an $h$-$^{11}$BN flake with a thickness of 40 nm. The scale bar is 400 nm. (b-d) PF-SNOM images at 1415 cm$^{-1}$ at tip-sample distances $d$ of 1, 2 and 5 nm respectively. (e) Averaged profiles of near-field signals along the orthogonal direction to the edge, revealing an increase of polariton wavelength $\lambda_p$ as $d$ increases. (f) Extracted $\omega$-$q_p$ dispersion relation of polaritons from the $h$-$^{11}$BN edge (blue squares, red circles, and green triangles) overlaid with the calculated dispersion relation (brown curve). $q_p$ exhibits distance-dependence at all frequencies. No preferred values of polariton momentum are observed.

**Formation of Geometric Resonances**

The comparison of the dispersion relations between the $h$-$^{11}$BN micro-disk and the reflective edge confirms the above-mentioned hypothesis and suggests that the momentum of polaritons in the circular micro-disk is quantized. The quantized spatial frequencies are determined by geometrically-formed standing wave modes when hyperbolic phonon polaritons propagate and reflect inside the micro-disk. The standing wave solutions can be analytically solved by a method described in previous literature on *h*-BN nano-resonators, by using Equation (1) [29] (see Supplementary Note 1 for the derivation):

$$J'_s(k_{sn}r_0 + \phi) = 0 \qquad (1)$$

where $J_S$ is the Bessel function of $s^{th}$ order, $r_0$ is the disk radius, $k_{sn}$ is the $n^{th}$ root to Equation (1) and the spatial frequencies of the standing wave, $n$ = 1, 2, 3…and $\phi_0$ is the anomalous phase shift at the edge, which is found to be -0.28 $\pi$ from fitting.

The standing wave solutions from Equation (1) have well-defined spatial frequencies ($k_{sn}$). When the tip launches polaritons with momenta that coincide with the characteristic frequency $k_{sn}$, one geometrical resonant condition is formed. Consequently, the amplitude of the tip-launched polaritons is amplified at these resonant conditions with several discrete spatial frequencies. $k_{sn}$ from standing wave solutions are marked as vertical dashed lines in Fig. 2e-f, from which we can clearly see that the fixed values of the polariton momentum are aligned with these discrete values of $k_{sn}$. Within the investigated frequency range, we observed three resonant frequencies: 1410, 1420 and 1428 cm$^{-1}$.

Also, when the metallic AFM tip is positioned at the crests of the standing waves, the coupling between the EM near-field of the tip and the polaritonic mode is amplified, leading to an increase of the scattering signals that is measurable by PF-SNOM. Consequently, the spatial fringes from PF-SNOM exhibit similar patterns as the amplitude square of the standing wave $\rho^2 = J_s^2(k_{sn}r_0 + \phi)$. Fig. 4 displays the PF-SNOM images and simulations based on the standing wave solution $\rho$. Simulation details are described in Supplementary Note 1. In comparison, the reflective $h$-$^{11}$BN edge of a large flake does not support geometric-dependent standing wave modes. Polaritons launched by the metallic AFM tip are not amplified at particular spatial frequencies (Fig. 3f). As a result, the dispersion relationship of the reflective edge lacks the quantization of the momentum and does not exhibit the characteristic zig-zag patterns in the dispersion relations of the $h$-$^{11}$BN micro-disk in Fig. 2f.

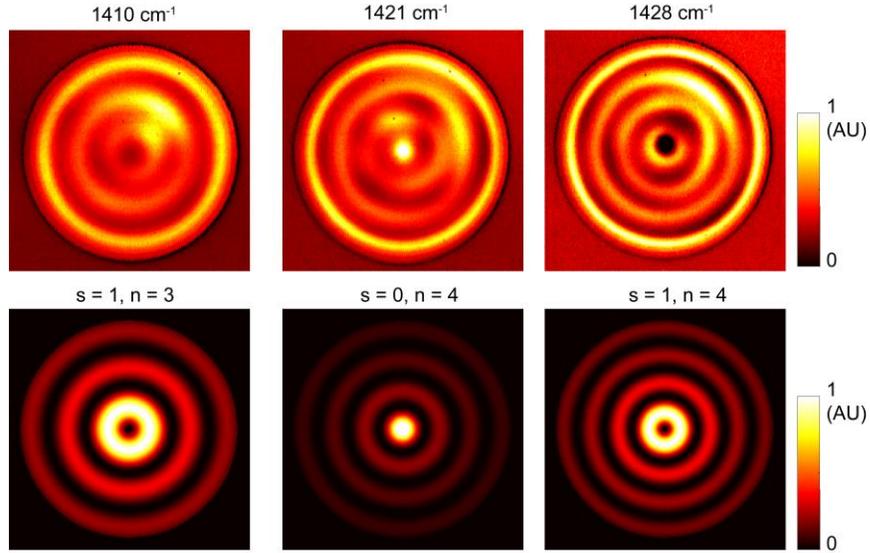

**Figure 4.** PF-SNOM images of the $h$-$^{11}$BN micro-disk at $d = 1$ nm at different resonant infrared frequencies (upper row) and their corresponding simulations of $\rho^2$ calculated from the standing wave solutions (lower row).

Changing the tip-sample distance can slightly tune the spatial frequency of PhP waves under conditions that lack geometrical resonance, such as in cases of the simple reflective edge and between two resonant conditions of the micro-disk (see Supplementary Fig. S4). The tuning of the spatial frequencies can be understood by the spatial distribution of the EM near-field underneath the tip at different tip-sample distances: the tip-sample configuration with a large tip-sample distance provides a loosely enhanced EM field with low spatial frequencies; the configuration with a small tip-sample distance strongly confines the EM field that spans high spatial frequencies (Fig.

1a-b). When the tip-sample distance is small, the PhP wave with a higher spatial frequency is excited more; when the tip-sample distance is large, the PhP wave with a lower spatial frequency is more present. On the other hand, under geometrical resonant conditions of the circular micro-disk, discrete spatial frequencies of the standing wave solutions are amplified. The small tuning of the spatial frequencies from different tip-sample distances is overshadowed by the dominant spatial frequency from the standing wave solutions. When the excitation of wavelength the PhP is between two geometric resonant conditions, the spatial frequency of the PhPs is not dominated by one standing wave, but from a superposition of two standing waves. Thus, the apparent momentum can again be tuned by the tip-sample distance. This behavior leads to the progression of the spatial fringe patterns as the tip-sample distance changes.

**Vertical Analysis of Near-field Response**

The 3D near-field response cube contains vertical sectional information on how the near-field response behaves when the tip-sample distance changes. Fig. 5 displays the vertical sectional images and analysis of the $h$-$^{11}$BN micro-disk. The green dashed line in Fig. 5a shows the position where the sectional profile was taken. Fig. 5b-c displays $S_{NF}(x, d)$ sectional image at infrared frequency of 1420 cm$^{-1}$ and 1425 cm$^{-1}$ respectively. A vertical signal decay range of less than 20 nm is observed, which is the dominant range of the near-field interaction between the tip and the sample. This observation can be understood by the fact that the phonon polaritons in $h$-$^{11}$BN are hyperbolic PhPs that travel in the volume of the material, thus the evanescent EM field outside $h$-$^{11}$BN is weak and not a part of the core component of hyperbolic PhPs inside $h$-$^{11}$BN.

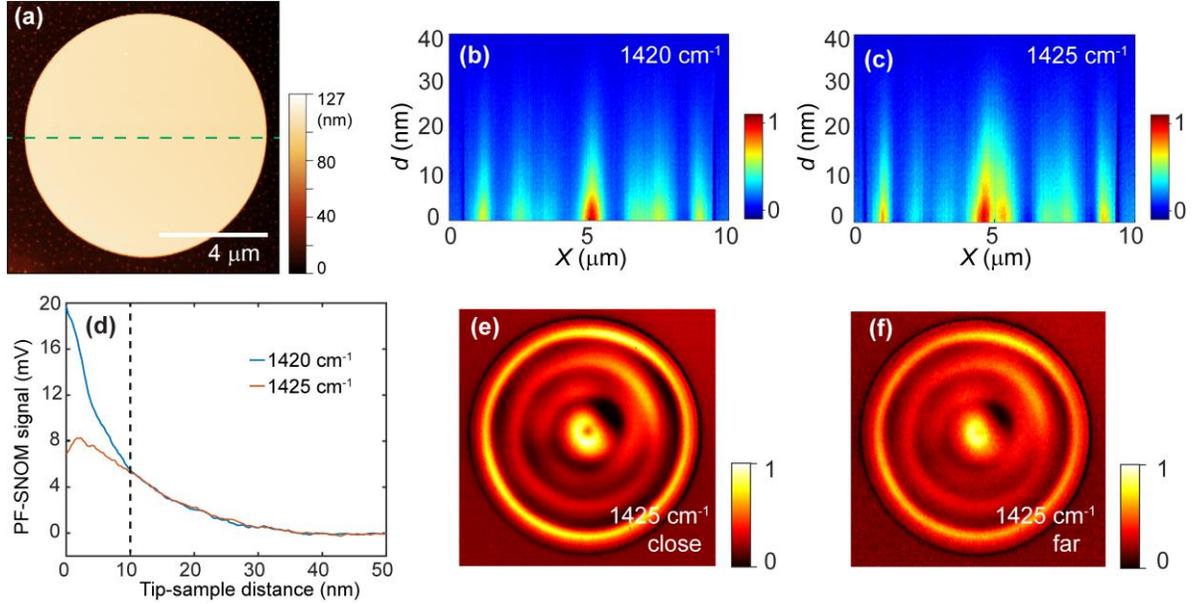

**Figure 5.** Vertical sectional near-field images and analysis. (a) AFM topography of the micro-disk. The green line across the center of the disk marks the position of vertical sectioning. (b-c) *x-d* sectional near-field responses at 1420 cm$^{-1}$ and at 1425 cm$^{-1}$, respectively. (d) Near-field vertical interaction curve at the center of the micro-disk at infrared frequencies of 1420 (blue) and 1425 cm$^{-1}$ (red). The black dashed line is the boundary between the close (d < 10 nm) and far range (d > 10 nm). Within the far range, two vertical behavior curves are effectively the same. (e) Reconstructed near-field image of 1425 cm$^{-1}$ at *d* = 0 nm, by extrapolating the exponential fitting of vertical response curves within the close range. Reconstructed image by fitting the data within the far range is shown in (f), in which the center dip disappears.

To examine the behavior of the vertical near-field interaction *in situ,* we read out the $S_{\text{NF}}(d)$ responses at the center of the micro-disk at 1420 and 1425 cm$^{-1}$, and display them in Fig. 5d. An interesting phenomenon appears: For the tip-sample distance region of d > 10 nm, two curves practically overlap, indicating a similar signal generation mechanism. However, at d < 10 nm, two vertical interaction curves exhibit drastically different behavior. The curve of 1420 cm$^{-1}$ exhibits a monotonic and rapid increase as *d* decreases; the curve of 1425 cm$^{-1}$ exhibit a slow increase as *d* decreases, at a short range of d = 2 nm, the signal even starts to decrease. How do we understand this behavioral difference? The infrared frequency of 1420 cm$^{-1}$ matches the geometrical standing wave mode of the polaritons in the *h*-$^{11}$BN micro disk, as revealed above; the infrared frequency of 1425 cm$^{-1}$ is located between two resonant standing wave modes. Under the resonant condition, the near-field tip probes the same polaritonic standing wave mode regardless of the tip-sample distance. When the infrared frequency is situated between two standing wave resonant modes, the near-field tip can moderately couple to both adjacent standing wave modes and excite a superposition of them.

At large *d*, the tip-enhanced electric field weakly couples relatively more of the standing wave mode with the lower momentum; at a small *d*, the strongly tip-enhanced electric field excites more of the standing wave mode with the higher momentum. As the centers of two adjacent standing wave modes of the 2D Bessel Function alternate between crest and trough (Fig. 4 and Supplementary Fig. S5), the mixture of two adjacent Bessel functions leads to destructive interference between the amplitudes of the polariton wave at the center of the micro-disk. As the tip-sample distance becomes smaller, the relative strength of the standing wave mode with high momentum increases, and the destructive interference become stronger, eventually leading to the dip of the vertical interaction curve as seen in Fig. 5d. The existence of such superposition is highlighted in Figs. 5e-f by reconstructing the near-field image at 1425 cm$^{-1}$ through extrapolating exponential fittings of vertical responses within close (0-10 nm) and far (10-50 nm) *d* ranges. The reconstructed image from the close range resembles the experimental tomographic image in Fig. 2c. However, in the reconstructed image from the far range, the center dip disappears. The difference between Figs. 5e and 5f is the proof of the destructive interference at the close *d* range. The non-trivial vertical decay behavior of polaritons is the manifestation of the quantization of momentum from the standing wave modes, and the shift of the mode superposition is due to the change of momentum coupling from the changing tip-sample distance.

**Discussion**

The advantage of PF-SNOM for near-field analysis over traditional s-SNOM analysis is the ability to access one additional vertical dimension along with the tip-sample distance $d$. In current s-SNOM, the spatial distribution of $x$-$y$ lateral near-field responses is collected as the nonlinearity of scattering signal over a range of tip-sample distance, and the progression of the lateral spatial distribution within the cantilever oscillation range is not accessible. As a result, the spectrum in the spatial frequency domain is averaged in the vertical range and inhomogeneously broadened. In contrast, PF-SNOM directly provides the vertical near-field response curve with well-defined tip-sample distances $d$ over a long range, which enables the collection of 3D near-field response cube for the near-field tomography to experimentally identify the geometrical resonances from both tomographic and sectional analysis. The addition of the near-field information in the vertical dimension enables the complete characterization of optical near-fields.

PF-SNOM and s-SNOM shares the same signal generation mechanism: the sharp metallic tip is used as both an antenna to generate a strong light confinement and a scatterer to emit detectable signals. Therefore, PF-SNOM can routinely provide nanometer resolution, which is better than other 3D near-field mapping methods based on aperture near-field scanning optical microscopy (NSOM), which uses the customized fiber as a light probe.[31-33] The effort towards characterizing vertical near-field response is also achieved by s-SNOM-based techniques with approach curve[34] or reconstruction method[35] at a single spot, but both methods are time consuming for obtaining tomographic images. Unlike these single spot methods, PF-SNOM can directly measure near-field responses in 3D space and provides a 3D data cube with rich information in only one scan, which improves the acquisition speed and enables verifications of numerical modeling on near-field responses. The ability to perform measurement on individual nanostructures also bypasses the inhomogeneous broadening of spectra due to spatial averaging.

The coupling between the launched phonon momentum and the momentum-quantized standing wave mode of $h$-$^{11}$BN structures could be useful for a wide range of applications that rely on both spectral and spatial selectivity. At the resonant frequency of the standing wave mode, the EM nearfield is strongly enhanced and spatially localized at certain positions of the disk, e.g., at 1420 cm$^{-1}$, the strongest nearfield is located at the disk center, yielding a near-field hotspot with a diameter of 500 nm, which suggests that the center of the $h$-$^{11}$BN micro-disk could be used for chemical sensing, as well as a platform with a strong field enhancement in mid-infrared systems or nonlinear optical mixing that involves mid infrared frequencies.[30, 36] To increase the spectral selectivity, the geometry of $h$-$^{11}$BN structures can be tailored to match resonances to infrared absorptions of investigated molecules. The mode quantization of the circular $h$-BN disk could also enable it serve as a momentum filter for selecting propagating PhPs with precisely defined momentum when integrated into a heterostructure.[37, 38]

**Conclusion**

In conclusion, we demonstrated the near-field analysis of $h$-$^{11}$BN microstructures based on the 3D near-field response cube from PF-SNOM with both tomographic and sectional analysis. The geometrical resonance from the standing wave solutions of the circular micro-disk leads to the quantization of the in-plane momentum of the polaritons. Between resonant conditions and for structures that lack geometric confinement, changing the tip-sample distance fine tunes the momentum of polariton. The 3D near-field tomography provides a convenient approach for *in situ* examination of polaritonic nano- and microstructures and is expected to be readily applicable to other two-dimensional materials and devices that have exquisite plasmonic and polaritonic

properties.[39-43]

**Methods**

**Materials**

The sample were prepared starting from a silicon wafer with 285 nm of thermal oxide. Optical lithography and gold evaporation were used to define alignment markers. Subsequently, the wafer was diced into chips. Isotopically pure *h*-$^{11}$BN crystal flakes were grown from a molten metal nickel-chromium solution using monoisotopic boron-11 as the boron source.[25] The *h*-$^{11}$BN was exfoliated using the Scotch tape technique and transferred on the substrate. The excess of glue was removed with acetone and IPA. E-beam lithography with ZEP resist was performed on the flakes and was followed by reactive ion etching with a fluorine-based recipe. Subsequently, ZEP was removed with remover PG, and the sample was rinsed with acetone and IPA.

**PF-SNOM setup**

The PF-SNOM setup used in this work is composed of an AFM (Bruker Multimode 8 with Nanoscope V controller with peak force tapping), a frequency tunable quantum cascade laser (QCL) (MIRcat, Daylight Photonics), a mercury cadmium telluride (MCT) infrared detector (KLD series, Kolmartech), and data acquisition devices (PXI-5122 and PXI-4461, National Instruments). Gold-coated AFM tips (HQ:NSC15/CR-AU, Mikromasch) were used. The infrared beam from QCL was focused on the tip apex with a gold-coated parabolic mirror (numerical aperture of 0.25) during experiments. The scattered light from the tip-sample region was collected by the same parabolic mirror, and then directed to the MCT detector. The light path of infrared laser was adjusted to maximize detected near-field signals. During the experiment, the AFM was operated in the peak force tapping mode of 4 kHz, where the tip is hold stationary and the peak-to-peak oscillation amplitude of the sample stage is set as 300 nm. Scattered signals from the MCT detector, cantilever deflection signals from the quadrant photodiode of AFM, and applied piezo voltages (x and y) from AFM are all recorded by two data acquisition devices simultaneously. In Figs. 1c-d, cantilever deflection and scattered signals from average of 200 cycles of peak force tapping were used. In all other PF-SNOM measurement, signals are from the average of 60 cycles. A custom script is used (MATLAB R2018b) to convert the raw data into a 3D response cube to get PF-SNOM images at different tip-sample distances, vertical near-field interaction curves, etc. The conversion between the time domain and the tip-sample distance domain and the algorithm to subtract far-field backgrounds are the same as described in the literature.[17]

**FDTD simulation**

FDTD simulations in Figs. 1a-b were conducted by Lumerical FDTD software (Lumerical Inc.). A gold cone with the end radius of 30 nm was used to simulate the AFM tip, and a Si disk with a radius of 4 μm and thickness of 80 nm was placed under the tip by a separation of 1 or 10 nm. The IR dielectric constant $\varepsilon = 11.7$ for Si was used.[44] The light source was set to be a p-polarized plane wave with an amplitude of 1 arbitrary unit and an incident angle of 15° to mimic the actual incident angle from the parabolic mirror. The incident angle is defined as the angle between the light (p-polarized in XZ plane) incident direction (which is also in XZ plane) and the horizontal plane (sample plane, or XY plane).

**Dispersion relation simulation**

The calculated dielectric function of $h$-$^{11}$BN in Fig. 2a is based on Equation 2.[3, 45]

$$\begin{cases} \varepsilon_\perp = \varepsilon_{\infty,\perp} + \varepsilon_{\infty,\perp} \dfrac{\omega_{LO,\perp}^2 - \omega_{TO,\perp}^2}{\omega_{TO,\perp}^2 - \omega^2 - i\omega\Gamma_\perp} \\ \varepsilon_\parallel = \varepsilon_{\infty,\parallel} + \varepsilon_{\infty,\parallel} \dfrac{\omega_{LO,\parallel}^2 - \omega_{TO,\parallel}^2}{\omega_{TO,\parallel}^2 - \omega^2 - i\omega\Gamma_\parallel} \end{cases} \quad (2)$$

where $\varepsilon_\perp$ and $\varepsilon_\parallel$ are in-plane ($x$ and $y$) and out-of-plane ($z$) dielectric functions; $\varepsilon_\infty$ is high-frequency permittivity; $\omega$ is the frequency of incident light; $\omega_{LO}$ and $\omega_{TO}$ are longitudinal and transverse optical phonon mode frequencies, and $\Gamma$ is the phonon damping. Parameters of $h$-$^{11}$BN are from Giles *et al.*[22], and are summarized in Table 1:

**Table 1. Fitting parameters of $h$-$^{11}$BN**

| $\varepsilon_{\infty,\perp}$ | $\omega_{LO,\perp}$ (cm$^{-1}$) | $\omega_{TO,\perp}$ (cm$^{-1}$) | $\Gamma_\perp$ (cm$^{-1}$) | $\varepsilon_{\infty,\parallel}$ | $\omega_{LO,\parallel}$ (cm$^{-1}$) | $\omega_{TO,\parallel}$ (cm$^{-1}$) | $\Gamma_\parallel$ (cm$^{-1}$) |
|---|---|---|---|---|---|---|---|
| 5.32 | 1608.7 | 1359.8 | 2.1 | 3.15 | 814 | 755 | 1 |

The $\omega - q_p$ dispersion relation (brown curves in Figs. 2e-f) is calculated based on the air/$h$-BN/SiO$_2$ structure according to Equation (3):[3, 39]

$$q_p = \frac{i\sqrt{\varepsilon_\parallel}}{z\sqrt{\varepsilon_\perp}} [\arctan(-i\varepsilon_a\sqrt{\varepsilon_\parallel \varepsilon_\perp}) + \arctan(-i\varepsilon_s\sqrt{\varepsilon_\parallel \varepsilon_\perp}) + \pi l] \quad (3)$$

where $z$, $\varepsilon_a$, $\varepsilon_s$ are the thickness of $h$-$^{11}$BN, dielectric functions of air and SiO$_2$ substrate, respectively. We use $z = 70$ nm, $\varepsilon_a = 1$, and $\varepsilon_s = 2.05$. The Si substrate used in this study generally has a 285 nm thick thermal oxide layer on top of the Si. This layer contains a mixture of $SiO_2$ and Si, the permittivities of which are very different in mid-infrared. At 1400 $cm^{-1}$, the relative permittivity of Si is 11.7,[44] while for $SiO_2$ it is 1.2.[46] This heterogeneous nature of substrate determines that we cannot simply use the permittivity of $SiO_2$. Therefore, we used a modified $\varepsilon_s$ = 2.05 (between 1.2 and 11.7) to match experimental results, which is consistent for additional data from another BN micro disk in supplementary Fig. 3.

**Conflicts of interest**

There are no conflicts to declare.

**Acknowledgments**

H.W. and X. G. X. acknowledge fruitful discussions and the help on the model simulation from Dr. Antonio Ambrosio, the help on the two-dimensional wave model and dispersion relation simulation from Dr. Michele Tamagnone and the help from Dr. William L. Wilson of Center for Nanoscale Systems of Harvard University for the fabrication of the BN micro-disk. X.G.X would like to thank Dr. Slava Rotkin and Dr. Gilbert Walker for helpful discussions. H. W. and X. G. X. thank for the support from Beckman Young Investigator Program. J. H. E. and J. L. appreciate the support from the National Science Foundation, award number CMMI 1538127.

# Supplementary Materials: Three-dimensional Near-field Analysis Through Peak Force Scattering-type Near-field Optical Microscopy

Haomin Wang[1], Jiahan Li[2], James H. Edgar[2], and Xiaoji G. Xu[1*]

[1]Department of Chemistry, Lehigh University, 6 E Packer Ave. Bethlehem, PA, 18015, USA

[2]Tim Taylor Department of Chemical Engineering, Kansas State University, Manhattan, Kansas 66506, USA

*Email: xgx214@lehigh.edu

**Supplementary Note 1**: Derivation of resonant conditions and calculation of fringe patterns

**Supplementary Figure S1**. s-SNOM and PF-SNOM images.

**Supplementary Figure S2**. Extraction of spatial frequencies.

**Supplementary Figure S3**. Additional PF-SNOM measurement on another $h$-$^{11}$BN micro-disk.

**Supplementary Figure S4**. Tuning $q_p$ by varying tip-sample distance $d$ through PF-SNOM.

**Supplementary Figure S5**. Fringe patterns at two adjacent resonant conditions.

**Supplementary Note 1**: Derivation of resonant conditions and calculation of fringe patterns

Resonant conditions of $h$-BN nano-disk resonators had been investigated thoroughly in the previous work.[1] According to it, the propagation of polaritons on the surface of $h$-BN micro-disk can also be described by a 2D wave equation:

$$(i\omega)^2 \rho = v_p^2(\omega)\nabla^2 \rho \tag{S1}$$

in which $\rho$ is the spatial charge density, $v_p$ is the phase velocity and $\omega$ is the angular frequency. Resonant conditions can be obtained by the standing wave solution of **Equation S1** using a Neumann boundary condition at the edge of disk with radius $r_0$, where we have:

$$\left.\frac{\partial \rho}{\partial r}\right|_{r=r_0} = 0 \tag{S2}$$

Solutions of the 2D wave equation are in forms of Bessel function with polar coordinate $(r, \theta)$:

$$\rho = J_s(kr)(A\cos(s\theta) + B\sin(s\theta))e^{i\omega t} \tag{S3}$$

in which $J_s$ is the Bessel function of $s^{\text{th}}$ order, $k$ is the wave vector of polariton and $A, B$ are arbitrary complex amplitudes. By solving **Equation S1** and **S2** together, one can get standing wave solutions that represents the resonant conditions of a circular disk:

$$\rho = J_s(k_{sn}r + \phi)(A\cos(s\theta) + B\sin(s\theta))e^{i\omega t}$$
$$\text{with } k_{sn} = n^{\text{th}} \text{ root of } J_s'(kr_0 + \phi) = 0 \tag{S4}$$

Note that a correction phase factor $\phi$ is added to account for the extra phase shift of polariton waves that have been reflected by the disk edge. In our case, we found $\phi = -0.28\pi$ best fit our experimental results, which agrees well with $-0.28\pi$ [1] and $-0.25\pi$ [2,3] found in literatures.

Measuring $\rho$ at different non-resonant and resonant conditions with an AFM tip and illumination light results to different signal contributions. The majority of PF-SNOM signal is from the so-called roundtrip component, which is circularly symmetric and independent of $\theta$.[1] Thus, we can estimate PF-SNOM signal as roundtrip component:

$$\text{PF-SNOM signal} \propto \rho^2 = J_s^2(k_{sn}r + \phi) \tag{S5}$$

which can be used to calculate fringe patterns that we've seen in PF-SNOM images. Since the anomalous phase shift $\phi$ affects most the outermost fringe and cannot be easily accounted due to the presence of fringing field,[1] in our simulation of $\rho^2$ we omit this term to achieve optimal agreement between simulated fringes and PF-SNOM fringes.

Solving **Equation S4** gives out a series of $k_{sn}$, which should be relevant to experimentally observed resonant conditions. To assign a proper $(s, n)$ pair to each resonant condition, we compare $k_{sn}$ with $q_p$ from FFT on PF-SNOM images at resonant conditions. The workflow is shown below:

1. Obtain resonant $q_p$ from experimentally observed PF-SNOM images. In our case, we chose

PF-SNOM images at 1410, 1421, and 1428 cm$^{-1}$ as three resonant conditions (since the momentum is locked no matter the change of tip-sample distance). Experimental fringe patterns at each resonant condition were extracted by performing a radial averaging from the disk center on each of resonant PF-SNOM image. Then, FFT was performed on each fringe pattern to get spatial frequency of the fringe $k_f$, which then was converted to the momentum of polariton $q_p$ by $q_p = \pi k_f$.

2. A series of $k_{sn}$ from solutions of **Equation S4** were plugged back into **Equation S5** to get simulated fringe patterns $\rho^2$.
3. Compare $k_{sn}$ with $q_p$ and $\rho^2$ with experimental fringe patterns. If they match, assign corresponding mode numbers $(s, n)$ to this resonant frequency.

Parameters that are used in these steps for resonant conditions shown in the main text **Figs. 2e-f** are summarized in **Table S1**, from which we found that the resonances that we observed by PF-SNOM mainly correspond to modes with $s = 0$ or $1$. $s = 0$ corresponds to modes that show the maximal signal at disk center, while $s = 1$ corresponds to modes that show the minimal at disk center (**Supplementary Fig. S5**).

**Table S1** Parameters used in the assignment of $(s, n)$

| Resonant frequencies (cm$^{-1}$) | $q_p$ ($\mu m^{-1}$) | $(s, n)$ | $k_{sn}$ ($\mu m^{-1}$) | Parameters used in solving **Eq. S4** |
|---|---|---|---|---|
| 1410 | 2.11 | (1,3) | 2.08 | $\phi = -0.28\pi$ <br> $r_0 = 4.5\ \mu m$ |
| 1421 | 2.48 | (0,4) | 2.44 | |
| 1428 | 2.84 | (1,4) | 2.78 | |

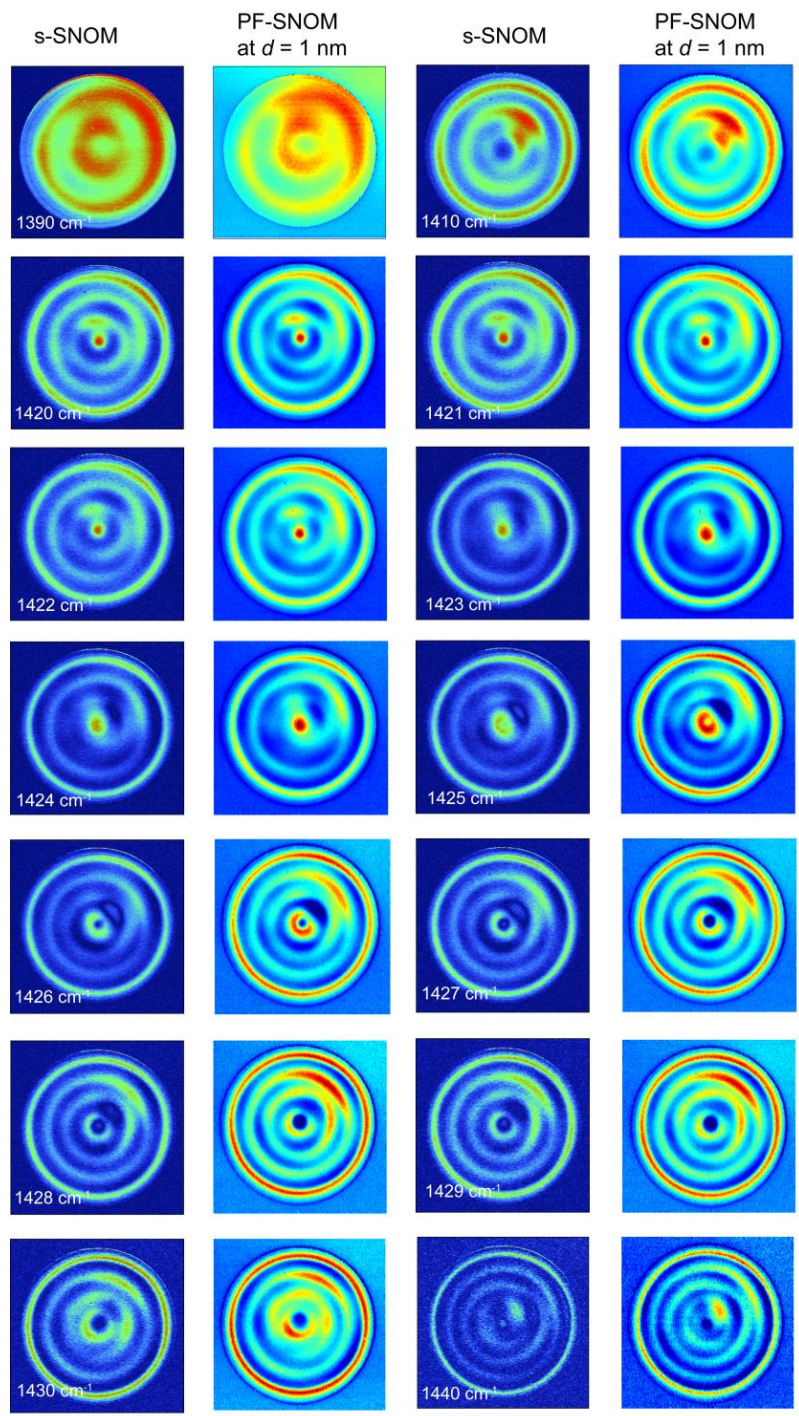

**Supplementary Figure S1**. s-SNOM and PF-SNOM images. s-SNOM images are obtained using the same homodyne setup as PF-SNOM. The AFM tip is mechanically driven by a piezo at its mechanical resonant frequency of 220 kHz. Scattered signals from the detector are routed into a lock-in amplifier, which demodulates the signal with the reference frequency of 220 kHz, the lock-in amplitude of 4[th] harmonic (880 kHz) is used to produce the near-field image.

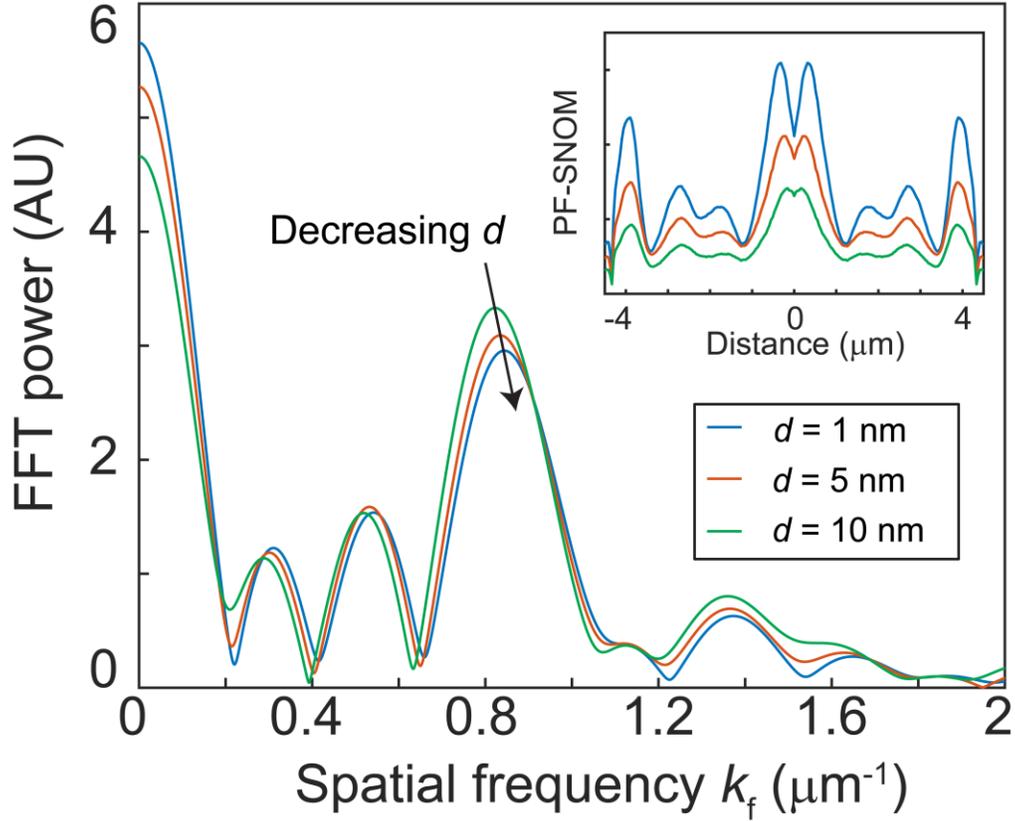

**Supplementary Figure S2**. Extraction of spatial frequencies. Fringe patterns at $\omega = 1425$ cm$^{-1}$ and $d = 1, 5, 10$ nm are extracted through radial averaging from the center of micro-disk and are displayed in the inset. The same procedure is applied to PF-SNOM images at other $\omega$ and $d$. Resulting fringe patterns are then Fourier Transformed to obtain the spatial frequency of the fringe $k_f$, and to get the polariton momentum $q_p$ by $q_p = \pi k_f$.

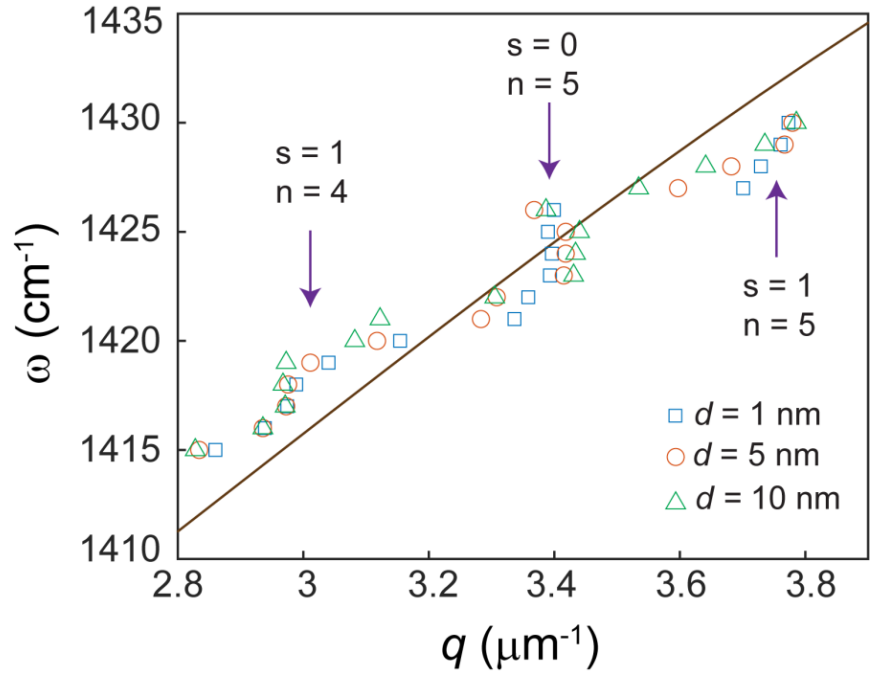

**Supplementary Figure S3**. Additional PF-SNOM measurement on another $h$-$^{11}$BN micro-disk. The thickness is 60 nm, and the diameter is 8.9 μm. Experimental data (blue squares, orange circles and green triangles) are overlaid with calculated dispersion relation (brown curve) and resonant conditions (vertical purple arrows, with assigned $(s, n)$ numbers). Similar to Figs. 2e-f in the main text, polariton momentum $q_p$ is less dependent of tip-sample distance $d$ at resonant conditions, and changes at different $d$ at non-resonant conditions.

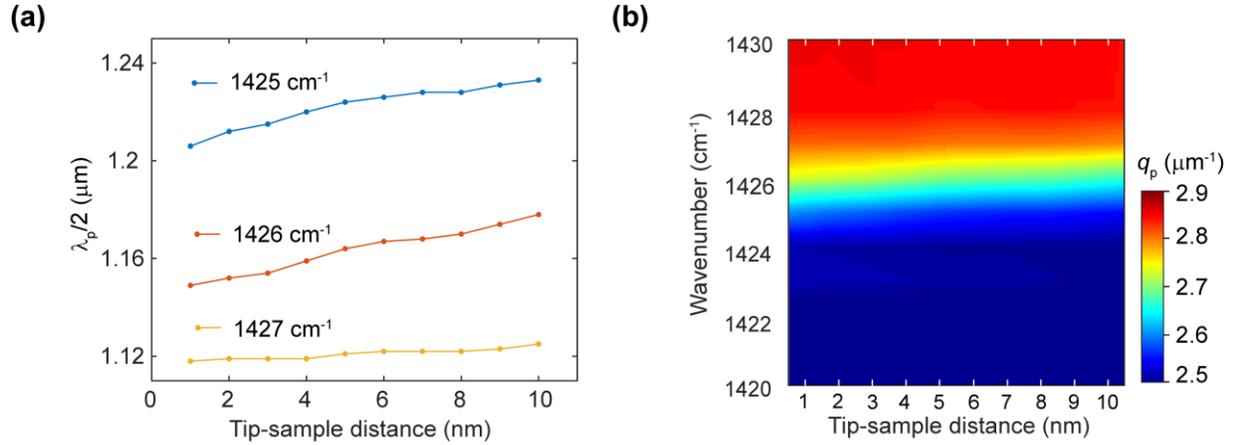

**Supplementary Figure S4**. Tuning $q_p$ by varying tip-sample distance $d$ through PF-SNOM. (a) Under non-resonant conditions such as $\omega = 1425, 1426$ and $1427$ cm$^{-1}$, the wavelength of the fringe pattern $\lambda_f$, polariton wavelength $\lambda_p = 2\lambda_f$ and polariton momentum $q_p = \frac{2\pi}{\lambda_p}$ can be tuned by changing tip-sample distance $d$. As this figure shows, as $d$ increases from 1 to 10 nm, $\lambda_f$ increases from 1.206, 1.149, 1.118 to 1.233, 1.178 and 1.125 μm for 1425, 1426 and 1427 cm$^{-1}$, respectively. (b) A false colormap of polariton momentum $q_p$ over $\omega = 1420-1430$ cm$^{-1}$ and $d = 1-10$ nm. At resonant conditions ($\omega = 1420 - 1422$ and $1428 - 1430$ cm$^{-1}$), $q_p$ does not change with $d$. On the contrary, at non-resonant conditions ($\omega = 1423 - 1426$ cm$^{-1}$), $q_p$ increases as $d$ increases. These results demonstrate that PF-SNOM is capable of tuning $q_p$ under non-resonant conditions and indicate that PF-SNOM can also be used to distinguish whether the system has extrinsic resonance.

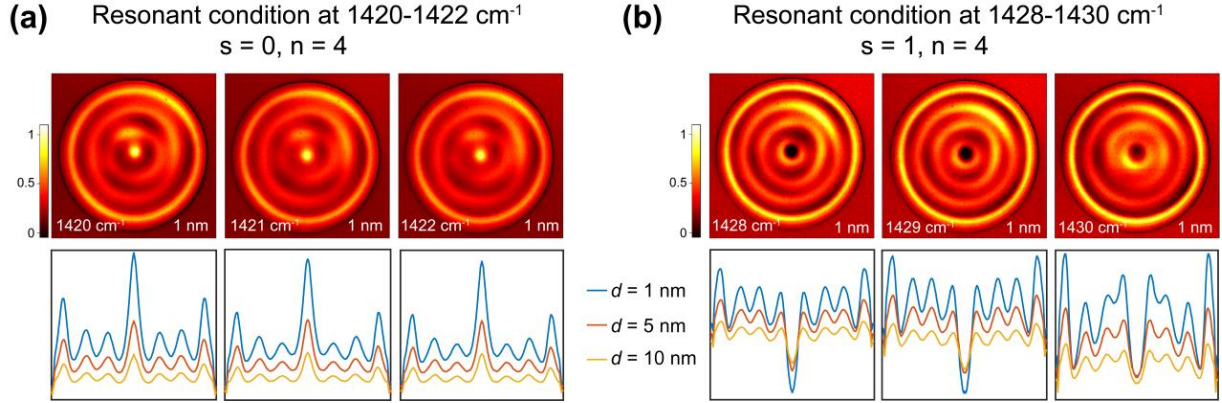

**Supplementary Figure S5**. Fringe patterns at two adjacent resonant conditions. According to PF-SNOM experimental results and model prediction (**Figs. 3b-c** in the main text), resonant momentum $q_\text{p}(s, n)$ increases with the increase of incident frequency $\omega$ with alternating $s$ between 0 and 1. In addition, the roundtrip component that we observed in PF-SNOM can be approximated as a square of the Bessel function (**Equation S5**). If we plug $r = 0$ (disk center) into **Equation S5** and omit the anomalous phase shift $\phi$, we will get $\rho^2 = J_s^2(0)$, which are constants determined simply by the order $s$ of Bessel function. With $s = 0$ and 1 alternating, we see PF-SNOM signal at the disk center alternating from bright ($s = 0$, $J_0^2(0)$ is local maximum) to dark ($s = 1$, $J_1^2(0)$ is local minimum).